# Predicting Residential Property Value in Catonsville, Maryland: A Comparison of Multiple Regression Techniques

Lee Whieldon
University of Maryland, Baltimore County
Catonsville, MD, USA
lewhiel1@umbc.edu

Huthaifa I. Ashqar
University of Maryland, Baltimore County
Booz Allen Hamilton
Washington, D.C., USA
hiashqar@vt.edu

# Abstract

This study aims to predict the prices of residential properties in Catonsville, MD based on publicly available tax assessment data maintained by Maryland government institutions. Recent legislation in Maryland has made it a requirement for local governments to provide up-to-date, easily accessible data to its constituents. This study explores predicting residential property prices in Catonsville, MD using three regression techniques: Linear, Ridge, and Lasso regression. Catonsville, MD was chosen as a case study since it is gaining a lot of interest as a suburb to Baltimore City, MD and is considered one of the most competitive markets for house sales. Extracting over 11,000 property records from Maryland's Open Data Portal (ODP), we transformed the data locally and applied regression techniques to predict the price. We used various independent features to predict the price of residential properties including prior year housing sales price, size of house, house age, street address type (single family or townhouse), if the house has a basement, dwelling type (center unit, end unit, split level, or standard unit), number of stories, and dwelling grade (scale from 1 to 6). Outperforming Linear and Ridge regression, Lasso regression can enhance the predictability of housing prices and significantly contribute to the correct evaluation of real estate price using only two years of historical data and without additional socioeconomic variables. We also found that among others; prior year housing sales price (positive), size of house (negative), and house age (positive) are the most significant factors in predicting housing sales price in Catonsville, MD.

# Introduction

The ability to accurately assess housing prices are vital to prospective homeowners, developers, real estate agents, and investors. The typical home buying process is complex, whereby it is difficult to understand all factors that affect house prices in a given market. Often, buyers are heavily reliant on real estate agents to help determine positive house factors and buying decisions during this event [1].

Recent legislation in Maryland has made it a requirement for local governments to provide up to date, easily accessible data to its constituents [2]; in Maryland's case, they have partnered with Socrata to maintain a single repository, called Maryland's Open Data Portal (ODP), allowing consumers to easily search for public datasets of their choosing [3]. One purpose of this paper is to leverage Maryland's publicly available data on residential property tax assessments and planning details. The other purpose is to compare different regression techniques to predict residential property prices. The data was pulled directly from Maryland's ODP using Socrata's API, which is maintained on a monthly basis by Maryland's Department of Assessment and Taxation, as well as Maryland's Department of Planning [4].

Catonsville, MD was chosen for this study due to the recently gained popularity as a suburb to Baltimore, Maryland. In addition, it was chosen so we could use the attributes provided within this dataset to compare properties in the same area. Any larger geographical comparison outside of a suburb like Catonsville, MD would require additional socioeconomic data elements, which was out of the scope for this study. We are assuming this inference based on census data provided by the United States Census Bureau; whereby, a collection of different statistics related to population, education, economy, income, and poverty are reviewed to be the same level on Catonsville area [5].

Previous studies described that regression methods, which is used to predict housing prices, require both a target variable and independent features. Independent features are usually grouped into two classifications including continuous independent features and indicator (or categorical) independent features [7]. Multiple Linear Regression is often the modelling method of choice when predicting housing prices in a finite, local neighborhood [8]. It includes techniques to analyze features with a relationship that affects a single continuous variable (i.e. housing prices). Several studies described the relationship between the target variable and the independent features using different dataset to predict house price. Feature selection and regularization in conjunction with regression techniques were used to fine tune their proposed model [6-9].

Ridge regression and lasso regression are often used in comparison with linear regression due to their ability to perform variable selection, regularization, and multicollinearity, enhancing the prediction accuracy of the model [7, 10, 11]. Unlike linear regression where all variables are considered equal, lasso regression penalizes the effect of features along with minimizing the error between predicted target variables and actual target variables, while ridge regression is a technique that helps address data situations where two or more features are highly correlated to the target variable. However, ridge and lasso regression are typically used to avoid overfitting that could occur within standard linear models [12, 13].

# Dataset and Methods

We used the Maryland's Department of Information and Technology's Socrata's API to extract data locally. The data includes approximately 11,000 residential properties in Catonsville, MD that have been assessed by Maryland's Department of Assessment and Taxation for the 2020 and 2019 fiscal years. The original dataset maintained in the portal contains over 200 features, but we reduced the number of features pulling only the proposed target variables and features required for the models.

To better understand the data, the minimum, maximum, mean, median, and standard deviation is provided for the housing sales price target variable, which was grouped into the number of stories each home contains. Table 1 provides the data summary statistics, which gives the indication that the 2-2½ story houses have the largest price range. In addition, the mean and the median of the housing sales prices increase as the number of stories increase.

Table 1: Summary statistics of the housing sales price variable based on number of stories.

| No of Stories | Minimum | Maximum | Mean | Median | Standard Deviation |
|---|---|---|---|---|---|
| 1 | $ 91,300 | $ 822,367 | $ 259,367 | $ 244,900 | $ 74,066 |
| 1 1/2 | $ 79,900 | $ 743,600 | $ 290,913 | $ 275,567 | $ 84,632 |
| 2 | $ 56,800 | $ 1,096,600 | $ 292,155 | $ 268,250 | $ 113,387 |
| 2 1/2 | $ 152,433 | $ 1,575,300 | $ 418,183 | $ 403,867 | $ 138,682 |
| 3 | $ 199,067 | $ 897,500 | $ 450,586 | $ 401,233 | $ 152,499 |

In addition, box plots may provide a better comparison between the distribution of the housing sales price target variable, breaking out the distribution by number of stories as shown in Figure 1. Note also that Figure 1 includes the number of observations within each box plot for each category. Interestingly, most houses in the dataset have 2 stories (over 8000 houses). There are also extreme outliers with houses in the data that have 2 ½ stories, which we eliminated from the used dataset to avoid their effect on the regression.

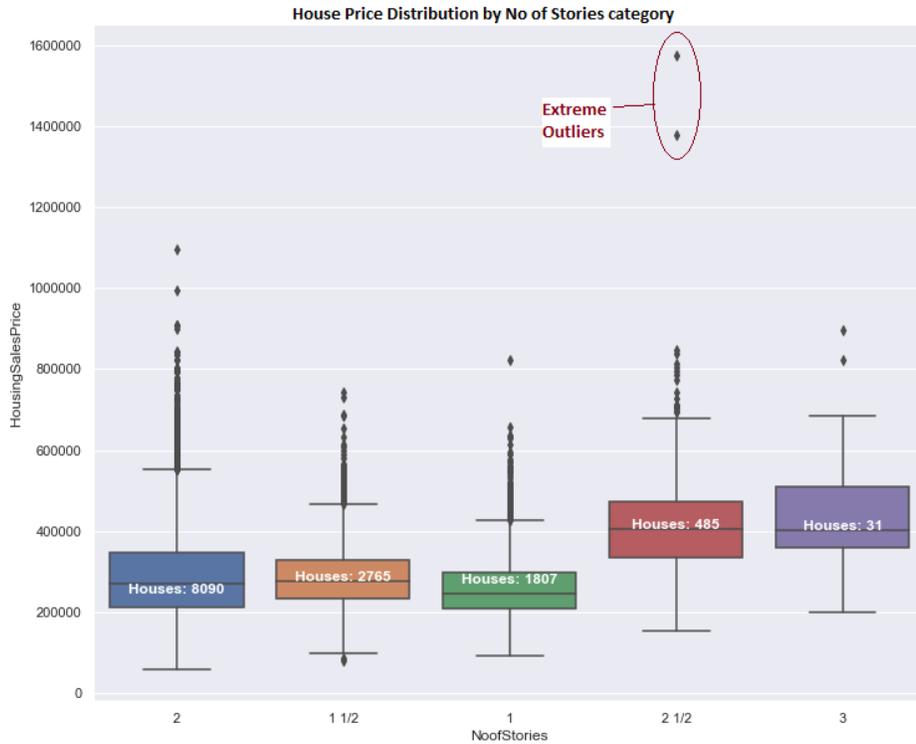

**Figure 1: Box plots, comparing the housing sales price for all houses grouped by the number of stories. Total number of houses by category is listed inside each box.**

Additional observations from the dataset can be noted using a histogram of the housing sales price, providing the frequency of residential properties by house price groups as shown in Figure 2. The majority of the residential properties fall in the $175k to $250k price range for the year of 2019.

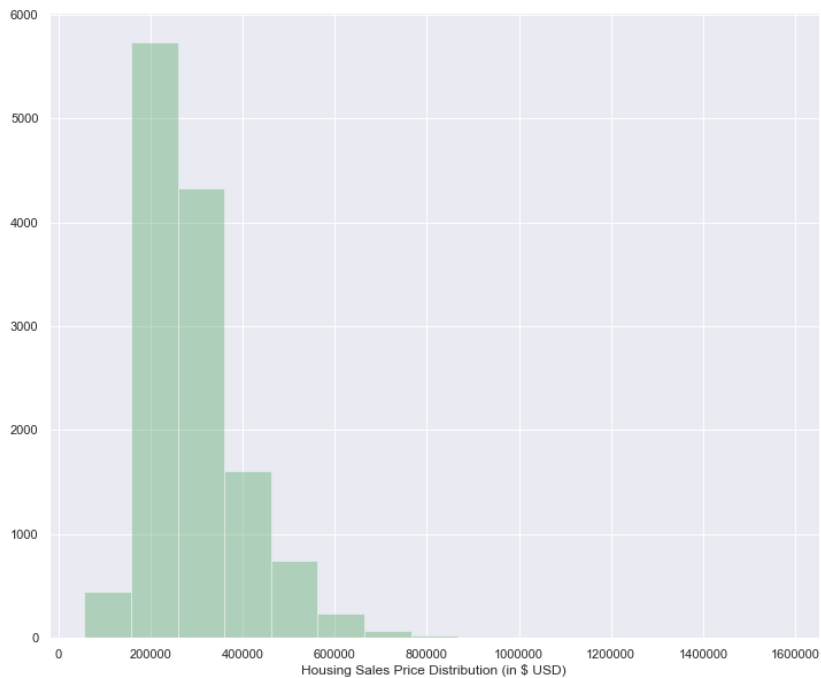

**Figure 2: Distributions of house sales price.**

## Linear, Ridge, and Lasso Regression

Since we are comparing the relationship between house sales price and its attributes, regression techniques are often the modelling tool of choice. Linear Regression is a standard in this type of analysis in that it assumes the relationship between a dependent, continuous variable, such as sales price, and dependent variables, like house attributes, is linear. In mathematical notation, if $\hat{y}$ is the predicted value [12-14]:

$$\hat{y}(\beta, x) = \beta_0 + \beta_1 x_1 + \cdots + \beta_i x_i$$

where $\beta_0$ is the intercept, $\beta = (\beta_1, \ldots, \beta_i)$ is the vector of the model coefficients, and $x = (x_1, \ldots, x_i)$ is the vector of the model predictors. Using Ordinary Least Squares, Linear Regression fits a linear model with coefficients $\beta = (\beta_1, \ldots, \beta_i)$ to minimize the residual sum of squares between the observed targets in the dataset, and the targets predicted by the linear approximation. Mathematically, it solves a problem of the form [12-14]:

$$\min_{\beta} ||X_\beta - y||_2^2$$

The coefficient estimates for Ordinary Least Squares rely on the independence of the features. When features are correlated and the columns of the design matrix $X$ have an approximate linear dependence, the design matrix becomes close to singular and as a result, the least-squares estimate becomes highly sensitive to random errors in the observed target, producing a large variance. This situation of multicollinearity can arise, for example, when data are collected without an experimental design. By analyzing the independent variables, one can then predict the value of the dependent variable, which in this case is house price. Moreover, Ridge and Lasso regression go even further than linear regression in that they adjust for multicollinearity within the data. Ridge and Lasso are well suited for housing datasets in that they help address bias in dataset itself [6, 10, 11, 15].

Ridge regression addresses some of the problems of Ordinary Least Squares by imposing a penalty on the size of the coefficients. The ridge coefficients minimize a penalized residual sum of squares [12-14]:

$$\min_{\beta} ||X_\beta - y||_2^2 + \alpha ||\beta||_2^2$$

The complexity parameter $\alpha \geq 0$ controls the amount of shrinkage; the larger the value of $\alpha$, the greater the amount of shrinkage and thus the coefficients become more robust to collinearity. So, ridge regression shrinks the coefficients and it helps to reduce the model complexity and multicollinearity [12-14].

On the other hand, Lasso is a linear model that estimates sparse coefficients. It is useful in some contexts due to its tendency to prefer solutions with fewer non-zero coefficients, effectively reducing the number of features upon which the given solution is dependent. For this reason, Lasso and its variants are fundamental to this study. Under certain conditions, it can recover the exact set of non-zero coefficients. Mathematically, it consists of a linear model with an added regularization term. The objective function to minimize is [12-14]:

$$\min_{\beta} \frac{1}{2n_{samples}} ||X_\beta - y||_2^2 + \alpha ||\beta||_1$$

The Lasso estimate thus solves the minimization of the least-squares penalty with $\alpha ||\beta||_1$ added, where $\alpha$ is a constant and $||\beta||_1$ is the $l_1 - norm$ of the coefficient vector. So, Lasso regression not only beneficial in reducing overfitting but also can be beneficial in feature selection [12-14].

## Analysis

This section breaks down the analysis into two work areas, demonstrating the steps performed to prepare and compare regression methods. All fields pulled from Socrata were 'object' data types originally and converted to the appropriate data types for the model (see Table 2).

Table 2: Features from MD Open Data source (renamed) and their type has also been provided to better understand how the data is leveraged.

| Field Name | Notes |
| --- | --- |
| Dwelling Type | Categorical Feature |
| Prior Year Housing Sales Price | Continuous Feature |
| Current Assessment Year | To be used to create 'House Age' continuous feature |
| Building Style Code and Description | To be used to create 'Has Basement' and 'No. of Stories' categorical features |
| Year Built | To be used to create 'House Age' continuous feature |
| Size of House | Continuous Feature |
| Street Address Type | Categorical Feature |
| Housing Sales Price | Target Variable |
| Dwelling Grade | Categorical Feature |

*Building Style Code and Description* field maintained two features including the *Number of Stories* and *Has Basement*; thus, we created two additional fields for the two features and appended them to the dataset. *Year Built* and *Current Assessment Year* fields were converted to integer data types and used to calculate the *House Age*. *Street Address Type* maintained two categories including *SF* for single family and *TH* for townhouse. Lastly, we converted *Housing Sales Price*, *Prior Year Housing Sales Price*, and *Size of House* fields from object to integer data types.

We also explored the correlation of each feature against the target variable (Housing Sales Price). The resulting correlation matrix contained 17 data attributes compared against one other, in which the closer to green the data attribute box color is assigned on the matrix represented a higher positive correlation to other data attributes in the dataset. However, the closer to red the data attribute box color is assigned on the matrix conversely represents a higher negative correlation. We used the correlations between the features to determine which of them will perform best in the developed model. We found that the *Prior Year Housing Sales Price* (correlation of 1.0), *Size of House* (correlation of 0.8), and *the Dwelling Grade* (correlation of 0.6) are the highest correlation to H*ousing*

*Sales Price* (as shown in Figure 3).

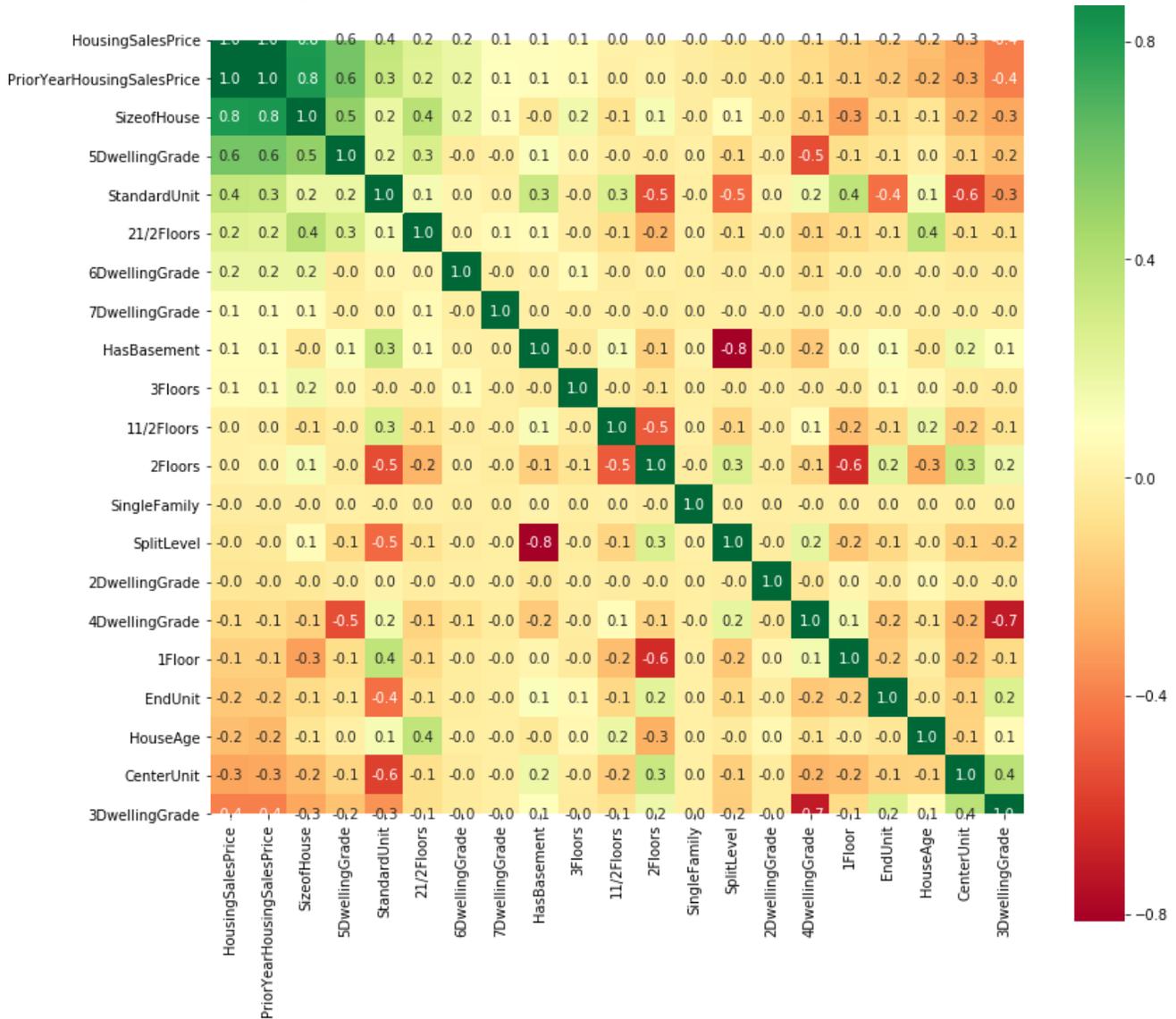

**Figure 3: Correlation matrix indicating that *Size of House* and *Dwelling Grade* features have the highest correlation to *Housing sales price*.**

To ensure the numerical features do not have a larger impact than other features in the regression models, the numerical features and the target variable were preprocessed using standardization. As an additional data exploratory step, the processed numerical features were reviewed and compared to the processed *Housing sales price* target variable (shown in Figure 4). As anticipated, *Size of House* and *Prior Year Housing sales price* are positively correlated to *Housing sales price*. On the other hand, *House Age* seems to have a little or no correlation with *Housing sales price.*

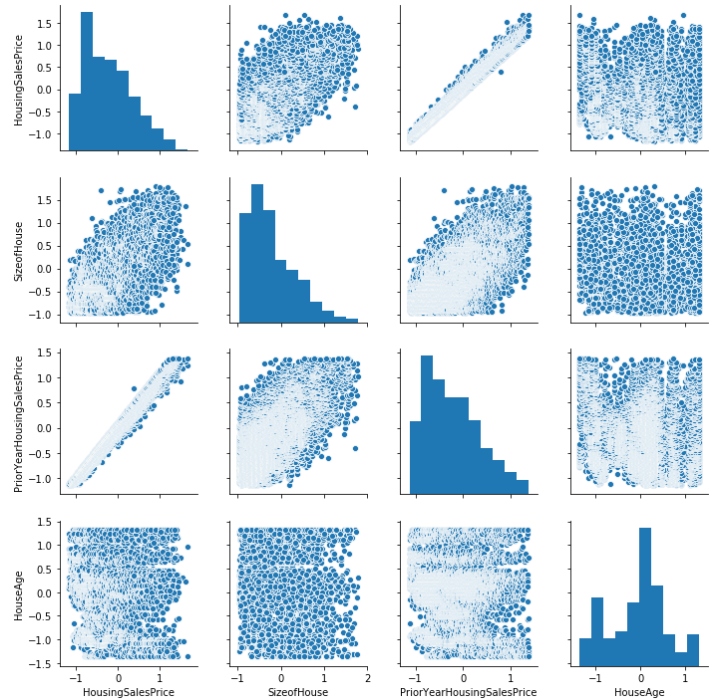

**Figure 4: Results of correlating Size of House, Prior Year Housing Sales Price, and House Age with Housing Sales Price, indicating positive correlation for the first two and no correlation with House Age.**

To better understand the correlation between the categorical features and the continuous target variable, Point Biserial Correlation was used [16]. This method can be used to measure the relationship between a binary variable (in this case, our dummy variables) and a continuous variable (*Housing sales price*). Return correlation values that are closer to -1 or +1 imply high correlation. Interestingly, the variables that represent the dwelling type of a residential property, if the house has a basement, the house with two and half floors, and the dwelling grade have a relatively high correlation with *Housing Sale Price* (see Table 3). Some of the results using this method was different from the previous one (see Figure 3) as the Point Biserial Correlation method is deemed to have a better performance in categorical variables [16]. In an hope to clear the confusion and test the performance of our methodology, we decided to include all the variables in the model. In the next section, we will discuss the results of the regression and attempt to link the inference to the results of this section.

**Table 3: Point Biserial Correlation scores for categorical features.**

| Feature Name | Correlation |
|---|---|
| Has Basement | 0.19 |
| Center Unit Dwelling Type | -0.31 |
| End Unit Dwelling Type | -0.20 |
| Split Level Dwelling Type | -0.01 |
| One Floor | -0.06 |
| One Half Floors | -0.07 |
| Two Floors | -0.04 |
| Two Half Floors | 0.14 |
| Three Floors | 0.01 |
| Single Family Address Type | 0.02 |
| Two Dwelling Grade | -0.02 |
| Three Dwelling Grade | -0.37 |
| Four Dwelling Grade | 0.11 |
| Five Dwelling Grade | 0.38 |
| Six Dwelling Grade | 0.00 |

## Results

The target and feature variables were split into training (70%) and testing (30%) datasets. To help prevent overfitting, we also used 7-fold cross validation for the training process of each developed model. To determine the accuracy of each model, multiple scoring parameters were used including Mean Squared Error and Adjusted $R^2$. The regression models were compared by reviewing the scoring parameters generated when fitting the models to the training dataset; this is performed by reviewing the average of the scores (Negative Mean Squared Error and Adjusted $R^2$) for each cross-fold validation training group. Linear, Ridge, and Lasso regressions were trained using the same training dataset and were verified using the same testing dataset as well. Table 4 shows the results of the regression including the estimates of regressors, standard deviation, $t$ test, and p-value for the Linear, Ridge, and Lasso regression models. Moreover, Figure 5 shows the fitted models against the testing dataset, whereas Table 5 shows the overall fit of the models in terms of score parameters.

In general, the values for each scoring parameter indicated that the three models performed at almost the same success rate when only considering Adjusted $R^2$. However, Lasso regression outperformed the Linear and Ridge regression when considering Negative Root Mean Squared. To understand how well the models performed on the testing data, the actual Housing sales price testing data was compared to the model's predicted value of Housing sales price. The difference was plotted between the actual versus predicted value, visualizing a univariate distribution for each record in the testing dataset. Linear and ridge regression has a near identical bimodal distribution; whereas, lasso regression has a normal distribution (see Figure 5).

The estimates of the regressors depict that the prior year housing sales price, size of house, if the house has basement, house age, whether the house is a center or end unit, the split level, and the number of floors are significant variables for predicting the house sales price. Three variables were found to be significant for the three models including prior year housing sales price, size of house, and house age, while the other variables seem to fluctuate between the different models.

Table 4: Estimates of regressors of the Linear, Ridge, and Lasso regression models.

| Regression | Linear | | | | Ridge | | | | Lasso | | | |
|---|---|---|---|---|---|---|---|---|---|---|---|---|
| Variables | Coef. | Std.E. | $t$ | $p$ | Coef. | Std.E. | $t$ | $p$ | Coef. | Std.E. | $t$ | $p$ |
| Intercept | 0.015 | 0.006 | 2.575 | <0.001 | 0.015 | 0.006 | 2.579 | 0.010 | -0.005 | 0.006 | -0.906 | 0.365 |
| Prior Year Housing Sales Price | 1.027 | 0.002 | 485.145 | <0.001 | 1.027 | 0.002 | 484.909 | <0.001 | 1.010 | 0.002 | 469.825 | <0.001 |
| Size of House | -0.021 | 0.002 | -9.056 | <0.001 | -0.021 | 0.002 | -9.048 | <0.001 | -0.013 | 0.002 | -5.522 | <0.001 |
| Has Basement | -0.028 | 0.004 | -6.210 | <0.001 | -0.028 | 0.004 | -6.205 | <0.001 | 0.006 | 0.005 | 1.341 | 0.180 |
| House Age | 0.036 | 0.001 | 26.243 | <0.001 | 0.036 | 0.001 | 26.230 | <0.001 | 0.035 | 0.001 | 25.126 | <0.001 |
| Center Unit | 0.005 | 0.003 | 1.416 | 0.157 | 0.003 | 0.003 | 0.949 | 0.342 | -0.007 | 0.003 | -2.424 | 0.015 |
| End Unit | -0.004 | 0.004 | -0.987 | 0.323 | -0.006 | 0.003 | -1.862 | 0.063 | -0.016 | 0.003 | -5.197 | <0.001 |
| Split Level | -0.018 | 0.004 | -4.350 | <0.001 | -0.020 | 0.004 | -5.190 | <0.001 | -0.014 | 0.004 | -3.499 | <0.001 |
| Standard Unit | 0.025 | 0.003 | 8.132 | <0.001 | 0.023 | 0.003 | 9.063 | <0.001 | 0.003 | 0.003 | 1.191 | 0.234 |
| 1 Floor | -0.001 | 0.004 | -0.323 | 0.747 | -0.003 | 0.003 | -1.010 | 0.312 | 0.000 | 0.003 | -0.139 | 0.889 |
| 1 and ½ Floors | -0.005 | 0.004 | -1.330 | 0.184 | -0.007 | 0.003 | -2.156 | 0.031 | -0.005 | 0.003 | -1.412 | 0.158 |
| 2 Floors | 0.003 | 0.003 | 0.838 | 0.402 | 0.001 | 0.003 | 0.226 | 0.821 | 0.000 | 0.003 | 0.107 | 0.915 |
| 2 and ½ Floors | 0.012 | 0.008 | 1.395 | 0.163 | 0.010 | 0.008 | 1.182 | 0.237 | 0.015 | 0.008 | 1.831 | 0.067 |
| 3 Floors | 0.000 | 0.000 | -1.070 | 0.285 | 0.000 | 0.000 | NaN | NaN | 0.000 | 0.000 | NaN | NaN |
| Single Family | 0.009 | 0.009 | 0.937 | 0.349 | 0.000 | 0.006 | 0.000 | 1.000 | 0.000 | 0.006 | 0.000 | 1.000 |
| 2 Dwelling Grade | -0.010 | 0.048 | -0.200 | 0.842 | -0.012 | 0.047 | -0.246 | 0.805 | -0.018 | 0.048 | -0.374 | 0.708 |
| 3 Dwelling Grade | 0.025 | 0.013 | 1.867 | 0.062 | 0.023 | 0.014 | 1.608 | 0.108 | 0.018 | 0.014 | 1.285 | 0.199 |
| 4 Dwelling Grade | -0.007 | 0.013 | -0.562 | 0.574 | -0.010 | 0.014 | -0.685 | 0.493 | -0.005 | 0.014 | -0.361 | 0.718 |
| 5 Dwelling Grade | 0.001 | 0.013 | 0.061 | 0.952 | -0.001 | 0.014 | -0.095 | 0.925 | 0.013 | 0.014 | 0.906 | 0.365 |
| 6 Dwelling Grade | 0.000 | 0.000 | NaN | NaN | 0.000 | 0.000 | NaN | NaN | 0.000 | 0.000 | NaN | NaN |

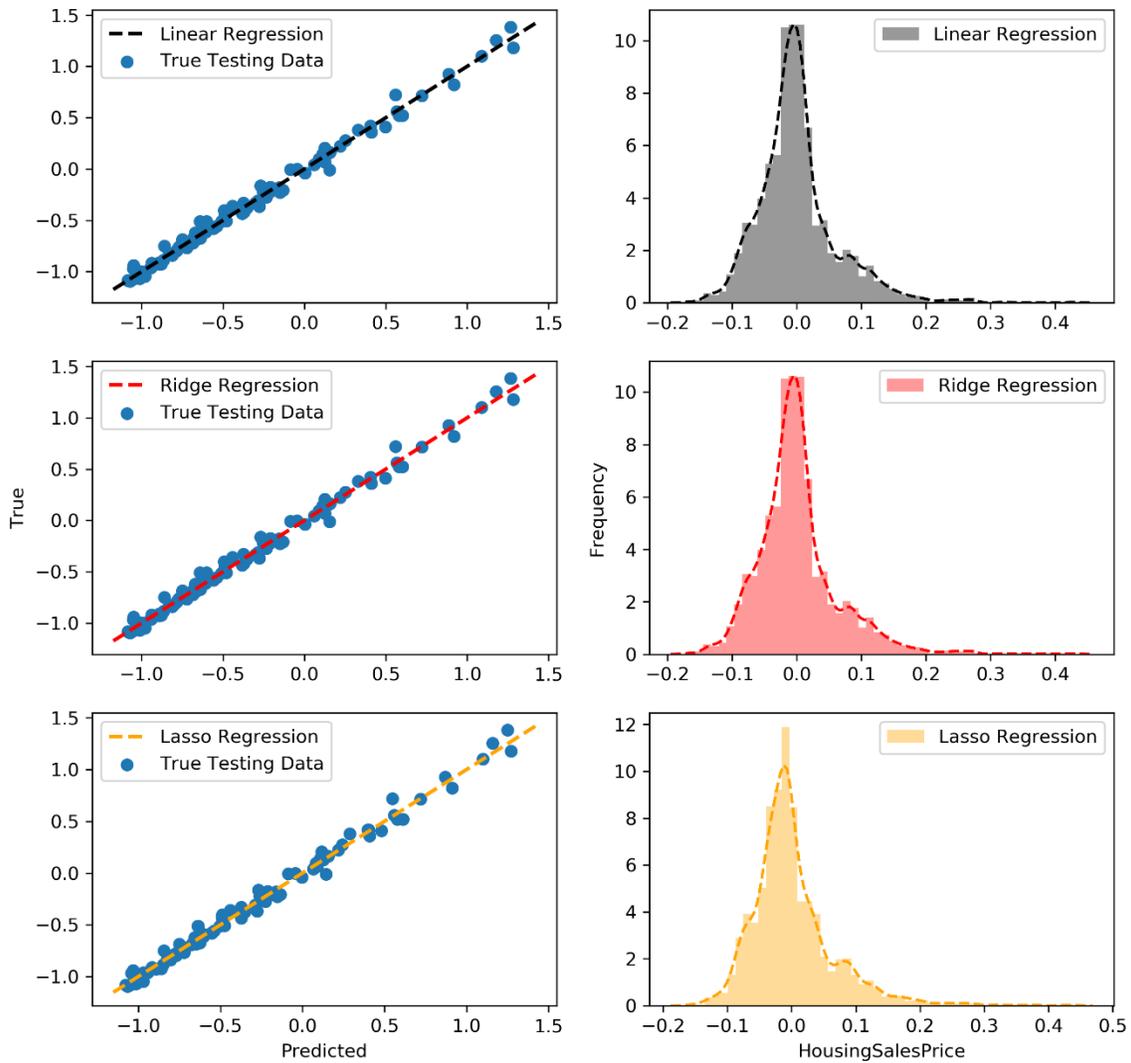

**Figure 5: Left: Linear, Ridge, and Lasso regression fitted against testing dataset. Right: Left skewed distribution of the actual house price to the predicted house price for each model**

**Table 5: Summary of the overall fit of the regression models using Adjusted $R^2$ and Negative Root Mean Squared.**

| Regression Method | Adjusted $R^2$ | Negative Root Mean Squared |
|---|---|---|

| Linear | 0.98 | $-6.6 \times 10^{19}$ |
| Ridge | 0.98 | $-0.0038$ |
| Lasso | 0.98 | $-0.13$ |

As Lasso regression is usually used to implement features selection, it seems to outperform the Linear and Ridge regressions and it includes the most important features to predict house sales prices. The prior year housing price is obviously one of the most important that has a positive relationship with the future house price. This was also found in many previous studies [17-19]. The size of the house, which may include lot area, masonry veneer area, floor area, garage car capacity, and green area has a negative relationship with the house sales price. The negative effect of the house size on the price has been discussed in many previous studies employing the envy effect or the reflected glory effect of housing consumption perspectives [20]. The study used a spatial autoregressive model to find that an increase in average house size of the eight nearest neighbors and the largest houses in six metropolitan statistical areas in Ohio has a negative effect on predicted house price. The house age in Catonsville, MD has a positive effect on the house sales price. It seems that, given that Catonsville housing market is very competitive with its architectural styles beyond the typical Colonial, there is a significant increase demand for houses that are relatively old as it could be more affordable [21]. Therefore, their house sale price has increased. As Maryland is a four-season state, central air conditioning and heating are necessary in a house. However, central and end air conditioning units are negatively related to the house prices. This might refer to the previously abovementioned trend of the increasing demand to older houses in Catonsville, which they usually do not have these units. It is thus hypothesized that people in Catonsville tend to buy old houses and renovate it as they see appropriate.

## Conclusion

Now that Maryland legislation has provided methods to access publicly available taxation and planning data, consumers can rely on regression methods to more accurately predict residential property prices in many areas including Catonsville, Maryland. By accessing the data through the Socrata API, transforming and analyzing the data using feature extraction, and building regression models, it can become a powerful tool to assist individuals when assessing a housing marketplace. Results show that Lasso regression outperformed Linear and Ridge regression using Maryland's public dataset in all the tests. Understanding that the public dataset be carefully reviewed and the appropriate features from the public dataset be selected, Lasso was sufficient to provide a fairly accurate prediction to the Catonsville, Maryland residential properties.

Previous studies pertinent to housing price predictions have focused on various methods including conventional statistical approaches that have some limitations of assumptions and estimations, and machine learning approaches that could be complex in building and interpreting, needs a huge amount of observations, and its training is costly. However, this study compares the performance of various easy-to-implement and interpretable methods of regression for a better and accurate housing price prediction. Our study shows that a Lasso and Ridge regressions can enhance the predictability of housing prices and significantly contribute to the correct evaluation of real estate price. Practically, mortgage lenders and financial institutions can employ these methods for better real estate property appraisal, risk analysis, and lending decisions. The potential benefits of using this model include reducing the cost of real estate property analysis, enabling faster mortgage loan decisions, and allow real estate owners and buyers to plan for future.

To further enhance the regression models covered in this study, it is recommended to combine other datasets related to the residential properties in Catonsville, MD, as well as external neighborhood features. Such examples of data to expand the scope of this model exploration include the location of schools and their National rankings, distance from public facilities likes parks and libraries, and socioeconomic factors like poverty and unemployment rates. Further iterations should consider incorporating other attributes of a house that are not included in the existing dataset, such as permits tracking renovations on the property, and other location features such as number of restaurants and commercial properties surrounding the property. Moreover, future research could examine further other geographical locations as locations is one of the most important factors in buying and selling real estate. Lastly, housing market can be influenced by macro-economic variables, which are not included in the proposed method. Future research should investigate the effect of macro-economic variables on housing price prediction. Nonetheless, this study could be deemed as a building block to use recently published data in accurately predicting housing sales price in various areas.